\documentclass[conference]{IEEEtran}
\usepackage{cite}
\usepackage{amsmath,amssymb,amsfonts}

\usepackage{algorithmic}
\usepackage{subfigure}

\usepackage{graphicx}
\graphicspath{{./figures/}}

\newcommand{\be}{\begin{equation}}
\newcommand{\ee}{\end{equation}}

\usepackage[bookmarks=true]{hyperref}
\hypersetup
{
    colorlinks=true, 
    linktoc=all,     
    linkcolor=black, 
    allcolors=black, 
    bookmarksopen=true,
    bookmarksnumbered=true,
    pdfstartview=, 
    pdfremotestartview=, 
}
\usepackage[all]{hypcap}
\usepackage{bookmark}
\usepackage{cleveref}

\def\BibTeX{{\rm B\kern-.05em{\sc i\kern-.025em b}\kern-.08em
    T\kern-.1667em\lower.7ex\hbox{E}\kern-.125emX}}
    
\begin{document}

\title{\huge{Novel Nonlinear Neural-Network Layers for High Performance and Generalization in Modulation-Recognition Applications}}

\author{\IEEEauthorblockN{John A. Snoap and Dimitrie C. Popescu}
\IEEEauthorblockA{ECE Department, Old Dominion University \\
Norfolk, VA 23529, USA \\
\{jsnoa001, dpopescu\}@odu.edu \vspace{-0.5cm} }
\and
\IEEEauthorblockN{Chad M. Spooner}
\IEEEauthorblockA{NorthWest Research Associates \\
Monterey, CA 93940, USA \\
cmspooner@nwra.com \vspace{-0.5cm} }
}

\maketitle

\begin{abstract}
The paper presents a novel type of capsule network (CAP) that uses custom-defined neural network (NN) layers for blind classification of digitally modulated
signals using their in-phase/quadrature (I/Q) components. The custom NN layers of the CAP are inspired by cyclostationary signal processing (CSP) techniques
and implement feature extraction capabilities that are akin to the calculation of the cyclic cumulant (CC) features employed in conventional CSP-based approaches
to blind modulation classification and signal identification. The classification performance and the generalization abilities of the proposed CAP are tested using
two distinct datasets that contain similar classes of digitally modulated signals but that have been generated independently, and numerical results obtained reveal
that the proposed CAP with novel NN feature extraction layers achieves high classification accuracy while also outperforming alternative deep learning (DL)-based
approaches for signal classification in terms of both classification accuracy and generalization abilities.
\end{abstract}


\section{Introduction}\label{sec:Intro}
Identification and classification of wirelessly transmitted digital communication signals into their respective modulation schemes without prior knowledge
of signal parameters is a challenging problem that occurs in both military and commercial applications such as spectrum monitoring, signal intelligence,
or electronic warfare \cite{Dobre_Sarnoff2005}, and conventional approaches to modulation classification employing either likelihood-based methods
\cite{Hameed_etal_TW2009} or CSP-based techniques~\cite{Spooner_Asilomar2000} can prove challenging \cite{Dobre_InstrumentMag2015}. As an
alternative, in recent years, DL-based techniques have been explored for blind classification of digitally modulated signals using for NN training and signal
recognition/classification either the raw I/Q signal components \cite{Latshaw_COMM2022, Snoap_CCNC_2022, Tim2018, Tim2017}, or pre-processing
them to extract CC features~\cite{Snoap_MILCOM2022} or other information that characterizes the signal amplitude/phase and frequency domain
representations \cite{Zhang2020, Kulin2018, Rajendran2018, Bu2020}.

We note that a common theme of the DL-based approaches proposed for signal classification involves the use of NN layers with structures developed
in the context of image processing, which appear to be poorly suited to extract information from time-domain signal samples. Specifically, the DL-based
approaches that use raw I/Q data as inputs along with conventional image processing NN layers for modulation classification have been shown to be unable
to generalize to new I/Q data when the distribution of the signal parameters of the new data differs slightly from the training data \cite{Latshaw_COMM2022,
Snoap_CCNC_2022}. Furthermore, pre-processing the I/Q data to obtain CCs and using these in conjunction with CAPs having conventional image
processing NN layers results in robust signal classification and generalization performance~\cite{Snoap_MILCOM2022}.

Motivated by the brittleness of DL-based classifiers that feed I/Q signal components into conventional image processing NN layers, and with the knowledge
that DL-based classifiers using CCs are robust, we propose a new approach to DL-based classification of digitally modulated signals, which involves custom-designed NN layers that perform specific nonlinear mathematical functions on the input I/Q data to enable downstream conventional image processing NN
capsules to identify specific features that are similar to those implied by the CCs.  By combining novel function layers that force the NN to extract periodic
(cyclic) features with capsules, our proposed approach enables a DL-based classifier to use the I/Q signal components as inputs, obtain generalization
that outperforms other types of NNs in the classification of digitally modulated signals \cite{Latshaw_COMM2022}, yet obviates the computational expense
of estimating CC features. We note that the robustness and generalization abilities of the proposed DL-based digital modulation classifier are assessed on
two distinct datasets that are publicly available \cite{IEEE_DataPort} and include signals with similar digital modulation schemes but which have been
generated independently using distinct signal distribution parameters, such that data from the testing dataset is not used in the training dataset. 

The remainder of this paper is organized as follows:  we provide related background information on CSP and the cycle frequency (CF) features relevant
for digital modulation classification in Section~\ref{sec:Background}, followed by presentation of the novel NN function layers and proposed NN structure
in Section~\ref{sec:Details}.  In Section~\ref{sec:Datasets} we provide details on the datasets used for training and testing the NNs, and we present numerical
 performance results in Section~\ref{sec:NumericalResults}.  We conclude the paper with discussion and final remarks in Section~\ref{sec:Conclusion}.

\section{CSP and Periodic Features for \\ Digital Modulation Classification}\label{sec:Background}
CSP provides a set of analytical tools for estimating distinct features that are exhibited by digital modulation schemes and that can be used to perform
classification of digitally modulated signals in various scenarios involving stationary noise and/or co-channel interference. These tools include higher-order
CC estimators \cite{Spooner_Asilomar2001, Spooner_Asilomar1995} and spectral correlation function (SCF) estimators \cite{Gardner_CSP_pt1_1994, Spooner_CSP_pt2_1994} that can be compared to a set of theoretical CCs or SCFs for classifying the digital modulation scheme embedded in a noisy signal.

To accurately estimate CC features for NN classification, specific signal parameters must be either known or accurately estimated from the I/Q data prior to
estimating the CC features~\cite{Snoap_MILCOM2022}. These parameters include the symbol interval, $T_0$, and symbol rate $1/T_0$, the carrier-frequency
offset (CFO), $f_0$, the excess bandwidth of the signal\footnote{For square-root raised-cosine (SRRC) pulse shaping this is implied by the roll-off parameter,
$\beta$.}, and the signal power level\footnote{This directly impacts the in-band signal-to-noise (SNR) ratio.}. These signal parameters define the CFs needed
for CC computation, with a key processing step involving blind estimation of the coarse second-order {\em CF pattern} for the signal. CFs $\alpha$ for which
a CC is non-zero for  typical communication signals include harmonics of the symbol rate $1/T_0$, multiples of the carrier frequency $f_0$,
and combinations of these two sets, such that $\alpha$ can be written as
\begin{equation}
\label{eq:basic_cf_psk_qam}
\alpha = (n-2m)f_0 \pm k/T_0.
\end{equation}
where $n$ is the order, $m$ is the number of conjugations, and $k$ is the set of non-negative integers typically constrained to a maximum value of $k=5$.

Although the set of possible CFs for all PSK/QAM/SQPSK signals is determined by the symbol rate and the carrier frequency offset
in~(\ref{eq:basic_cf_psk_qam}), most signals exhibit only a subset of the maximum possible set.  Moreover, there are only a few basic patterns:
BPSK-like, QAM-like, $\pi/4$-DQPSK-like,  8-PSK-like, and staggered QPSK (SQPSK)-like.  If we can find the pattern, we can also determine
the actual number of cycle frequencies needed to fully characterize the modulation type through its set of associated CC values.

Two distinct features that are meaningful for identifying which of the five basic CF patterns are present in a signal of interest are~\cite{Gardner_CSP_pt1_1994}:
\begin{enumerate}
\item The Fourier transform of the squared signal
\item The Fourier transform of the quadrupled signal
\end{enumerate}
Furthermore, features that are meaningful in relation to even-order CC estimates are:
\begin{enumerate}
\item The squared signal
\item The quadrupled signal
\item The signal raised to a power of six
\item The signal raised to a power of eight
\end{enumerate}
Since second and higher even orders in both the time- and frequency-domains are necessary to estimate CC features, we create custom NN function layers
that perform generic mathematical equations to provide feature extraction layers that are more consistently meaningful (in that they are proportional to CC
features) than the I/Q inputs prior to the trainable branches and final layers with many learnable hyperparameters. The aforementioned list of features is not
exhaustive in regards to CCs but only includes orders where no conjugations have been performed ($m = 0$).  This subset of CC features was chosen for
these proposed feature extraction layers to limit the size and complexity of the resulting NN structure.

\section{CAP with Custom Feature Extraction NN Layers for Modulation Classification}\label{sec:Details}
\begin{figure}[t]
\begin{center}
{\includegraphics[width=\linewidth]{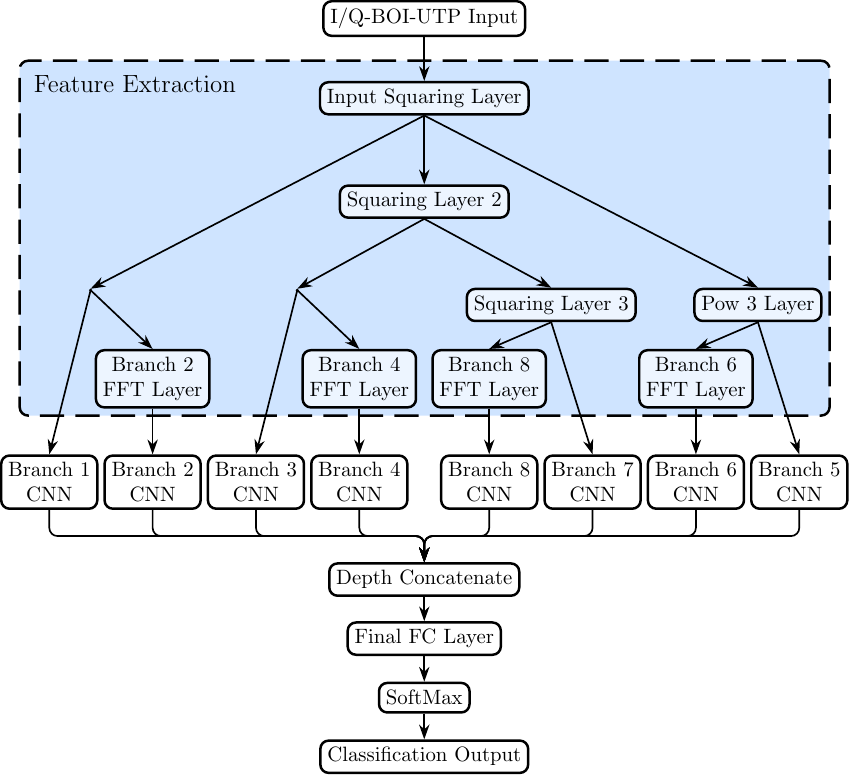}}
\vspace{-0.75cm}
\caption{CAP topology with novel custom NN layers performing feature extraction for each branch.}\label{fig:CAP_Topology}
\end{center}
\vspace{-0.65cm}
\end{figure}

To consistently extract the second, fourth, sixth, and eighth-order signal features in both time- and frequency-domain using the I/Q signal data, we only need
to implement three custom NN layers for the needed nonlinear computations: 
a squaring layer, a raise to the power of three (or Pow3) layer, and a fast Fourier
transform (FFT) layer. These custom layers can then be connected together as shown in Fig.~\ref{fig:CAP_Topology}, such that their outputs branch out until
all eight feature types have been extracted.

Noting that the I/Q signal data is input as pairs of real numbers $(I,Q)$, the squaring and power-of-three layers can be implemented as follows:

--~Squaring Layer:
\vspace{-0.2cm}
\begin{align}
I_{Output} &= \left( I \times I \right) - \left( Q \times Q \right) \\
Q_{Output} &= 2 \times I \times Q
\end{align}

\vspace{-0.2cm}
--~Pow3 Layer:
\vspace{-0.2cm}
\begin{align}
I_{Output} &= \left( I \times I \times I \right) - \left( 3 \times I \times Q \times Q \right) \\
Q_{Output} &= \left( 3 \times I \times I \times Q \right) - \left( Q \times Q \times Q \right)
\end{align}
\vspace{-0.5cm}

For the FFT layer, only the magnitude is needed at the output, with the zero frequency bin in the center, as this information is most likely to reveal
CF patterns present in the signal. Therefore, after taking the FFT, its absolute value is obtained to ensure the output magnitude is a real number,
suitable for use by downstream trainable NN layers.

\begin{table}
\centering
\caption{CNN Branch Layout}
{
\vspace{-0.25cm}
\begin{tabular}{ c c c }
\hline\hline
Layer & (\# Filters)[Filter Size] & Activations \\
\hline
Input			& 							& $32,768 \times Y$ \\
ConvMaxPool		& ($16$)[$23 \times Y$]	& $16,384 \times 16$ \\
ConvMaxPool		& ($24$)[$23 \times 16$]	& $8,192 \times 24$ \\
ConvMaxPool		& ($32$)[$23 \times 24$]	& $4,096 \times 32$ \\
ConvMaxPool		& ($48$)[$23 \times 32$]	& $2,048 \times 48$ \\
ConvMaxPool		& ($64$)[$23 \times 48$]	& $1,024 \times 64$ \\
ConvAvgPool		& ($96$)[$23 \times 64$]	& $96$ \\
FC				& 							& \# Classes \\
\hline\hline
\end{tabular}
\vspace{-0.25cm}
}\label{table:CNN_Branch}
\end{table}
\begin{table}
\centering
\caption{ConvMaxPool Layer}
{
\vspace{-0.25cm}
\begin{tabular}{ c c c c }
\hline\hline
Layer 		& (\# Filters)[Filt Size] & Stride & Activations \\
\hline
Input		& ($A$)[$B \times C$]	& 					& $X \times Y$ \\
Conv		& ($A$)[$B \times C$]	& [$1 \times 1$] 	& $X \times (Y \cdot A/C)$ \\
Batch Norm	& 						& 					& $X \times (Y \cdot A/C)$ \\
ReLU		& 						& 					& $X \times (Y \cdot A/C)$ \\
Max Pool	& ($1$)[$1 \times 2$]	& [$1 \times 2$]	& $(X/2) \times (Y \cdot A/C)$ \\
\hline\hline
\end{tabular}
\vspace{-0.25cm}
}\label{table:ConvMaxPoolLayer}
\end{table}
\begin{table}[h!]
\centering
\caption{ConvAvgPool Layer}
{
\vspace{-0.25cm}
\begin{tabular}{ c c c c }
\hline\hline
Layer 		& (\# Filters)[Filter Size] & Stride & Activations \\
\hline
Input		& ($A$)[$B \times C$]	& 					& $X \times Y$ \\
Conv		& ($A$)[$B \times C$]	& [$1 \times 1$] 	& $X \times (Y \cdot A/C)$ \\
Batch Norm	& 						& 					& $X \times (Y \cdot A/C)$ \\
ReLU		& 						& 					& $X \times (Y \cdot A/C)$ \\
Avg Pool	& ($1$)[$1 \times X$]	& [$1 \times 1$]	& $1 \times (Y \cdot A/C)$ \\
\hline\hline
\end{tabular}
\vspace{-0.5cm}
}\label{table:ConvAvgPoolLayer}
\end{table}

The CAP includes eight convolutional neural network (CNN) branches that implement the primary capsules and will be trained to classify eight
common digital modulation schemes: BPSK, QPSK, 8-PSK, $\pi/4$-DQPSK, MSK, 16-QAM, 64-QAM, and 256-QAM. The primary capsules
contain the majority of the network's learnable hyperparameters, with each branch having as its input one of the eight desired features.
At the output of each branch there is a fully connected layer with eight outputs, which is used to reduce the branch output size to the number of
modulation classes in the training dataset; this is done to ensure that each branch has the ability to distinguish between all modulation types in
the training dataset. This ability is important because, while not every branch will be able to fully distinguish between all modulation types, each
branch will be able to identify which classes it can distinguish between and this determination can be made during NN training. For example, due
to the nature of the CF patterns, we expect that the branch with second-order frequency-domain inputs will be able to distinguish between BPSK
and MSK, but all other modulation types will not be distinguishable on this branch. Likewise, the branch with fourth-order frequency-domain inputs
should be able to distinguish between 8-PSK-like, $\pi/4$-DQPSK-like, and SQPSK-like patterns, but QAM-like and BPSK-like CF patterns may
not be distinguishable to this branch.
Each branch, having opportunity to distinguish between all eight modulation types, then has its outputs concatenated together with the other branches followed by a final
fully connected layer.  This last fully connected layer learns the appropriate weights to apply to the outputs of each CNN branch so that each branch's
ability is combined together to obtain the minimum error on the training dataset.

The overall structure of the proposed CAP with custom feature extracting NN layers is shown in Fig.~\ref{fig:CAP_Topology}, with the characteristics of
the subsequent CNN branches for classifying digitally modulated signals outlined in Tables~\ref{table:CNN_Branch}~--~\ref{table:ConvAvgPoolLayer}.
The number of filters $A$ and filter size $\left[ B \times C \right]$ for each convolutional layer are defined in Table~\ref{table:CNN_Branch} as are the
number of output activations (e.g., $X \times Y$) for each layer.

\begin{table}[htbp]
\centering
\caption{Dataset Signal Generation Parameters}
{
\vspace{-0.25cm}
\begin{tabular}{ c c c }
\hline\hline
Parameter & \texttt{CSPB.ML.2018} & \texttt{CSPB.ML.2022} \\
\hline
Sampling Frequency, $f_s$ & 1 Hz & 1 Hz \\
CFO, $f_0$ & $U(-0.001, 0.001)$  & $U(0.01, 0.02)$ \\
Symbol Period, $T_0$, Range & $[1, 23]$ & $[1, 29]$ \\
Excess Bandwidth, $\beta$, Range & $[0.1, 1]$ & $[0.1, 1]$ \\
In-Band SNR Range (dB) & $[0, 12]$ & $[1, 18]$ \\
SNR Center of Mass & $9$ dB & $12$ dB \\
\hline\hline
\end{tabular}
}\label{table:SigGenParms}
\vspace{-0.5cm}
\end{table}

\section{Datasets Used for NN Training and Testing}\label{sec:Datasets}
To train and test the performance of the proposed CAP (including its generalization ability) we use two publicly available datasets that both contain signals
corresponding to the eight digital modulation schemes of interest (BPSK, QPSK, 8-PSK, $\pi/4$-DQPSK, MSK, 16-QAM, 64-QAM, and 256-QAM). The two
datasets are available from \cite{IEEE_DataPort} as \texttt{CSPB.ML.2018} and \texttt{CSPB.ML.2022}, and their signal generation parameters are listed
in Table~\ref{table:SigGenParms}.

We note that the CFO ranges corresponding to signals in the two datasets are non-intersecting, which allows evaluation of the generalization ability of
the trained CAP. Specifically, if the CAP trained on a large portion of the \texttt{CSPB.ML.2018} dataset displays high classification accuracy on the remaining
signals in the dataset, and its performance when classifying signals in the \texttt{CSPB.ML.2022} dataset is at similarly high levels, then the CAP is robust
and has a high ability to generalize\footnote{Likewise, if its classification accuracy on signals in \texttt{CSPB.ML.2018} is high, while on signals in
\texttt{CSPB.ML.2022} is low, then its generalization ability is also low.}.

Prior to training the proposed CAP on these datasets, a blind band-of-interest (BOI) detector~\cite{BOIdetector} is used to locate the bandwidth for
signals in the datasets, filter out-of-band noise, and center the I/Q data at zero frequency based on the CFO estimate provided by the BOI detector.
Finally, the I/Q data is normalized to unit total power (UTP) so that it does not prevent the activation functions of the NN from converging. To show that
using a blind BOI detector to shift the signal to an estimated zero frequency does not necessarily enable an I/Q-trained CAP such as the one
in~\cite{Latshaw_COMM2022} to generalize between the \texttt{CSPB.ML.2018} and \texttt{CSPB.ML.2022} datasets, we also use this data to
retrain the CAP in~\cite{Latshaw_COMM2022} as an alternative to provide a point of comparison with the proposed CAP.

\section{Network Training and Numerical Results}\label{sec:NumericalResults}
The proposed CAP and the alternative CAP in~\cite{Latshaw_COMM2022} have been implemented in MATLAB and trained on a high-performance
computing cluster, such that each CAP was trained and tested two separate times as follows:

\begin{figure}
\begin{center}
{\includegraphics[width=\linewidth]{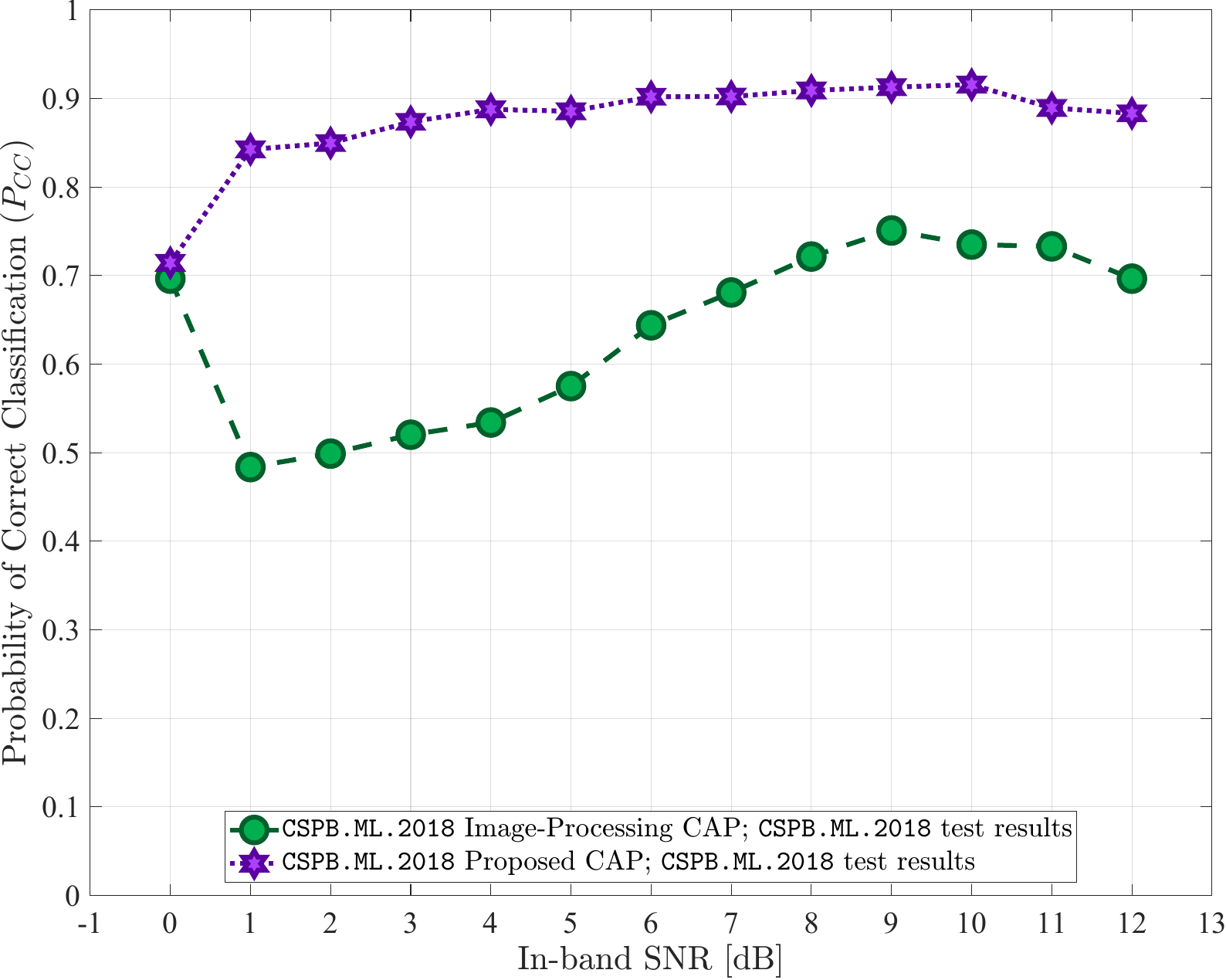}}
\caption{\vspace{-0.25cm}
Performance of CAPs trained and tested on \texttt{CSPB.ML.2018} dataset.}\label{fig:SNR_CTest_CTrained_All}
\end{center}
\end{figure}
\begin{figure}
\begin{center}
{\includegraphics[width=\linewidth]{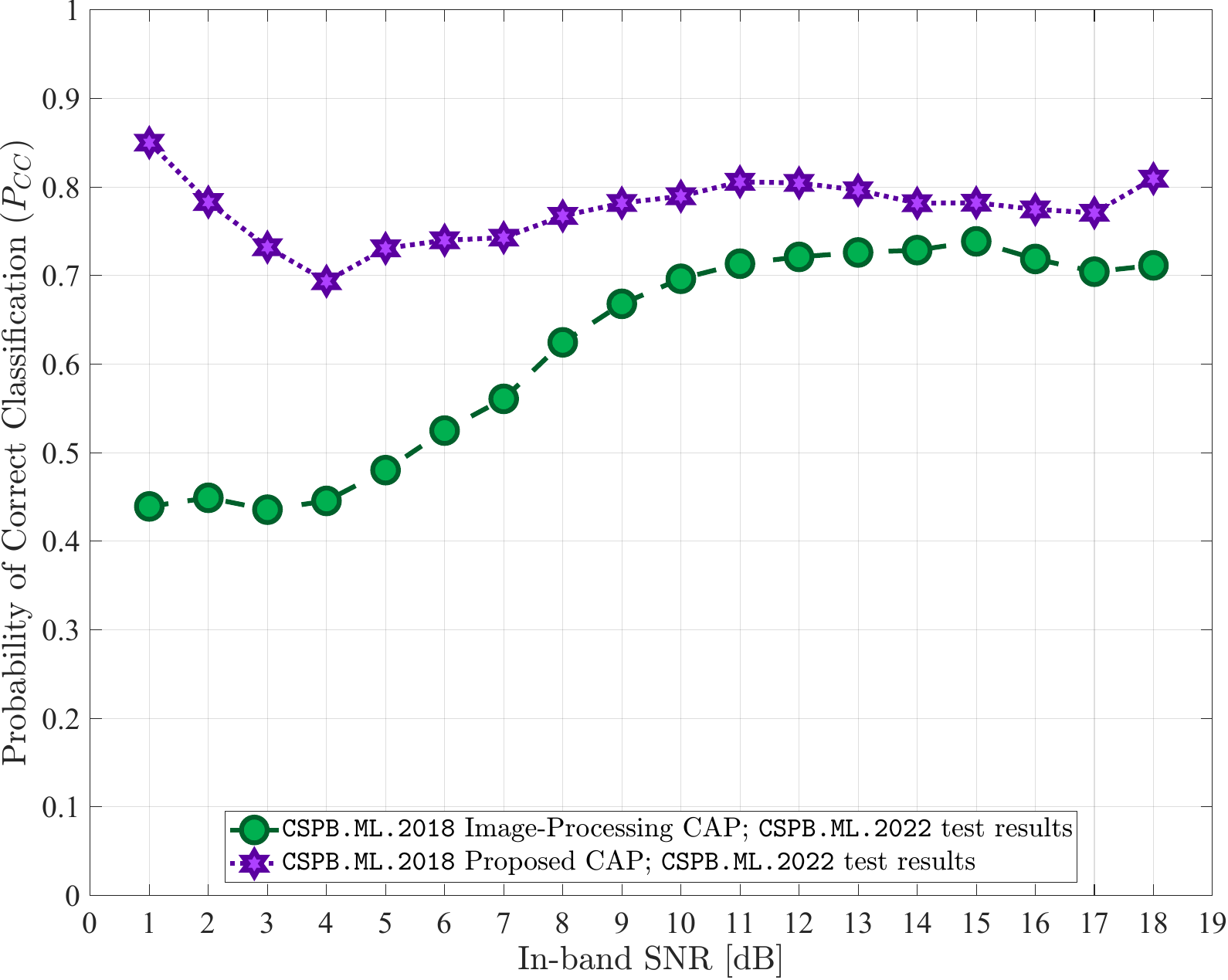}}
\caption{Generalization test results for CAPs trained on dataset \texttt{CSPB.ML.2018} and
tested on \texttt{CSPB.ML.2022}.}\label{fig:SNR_GCTest_CTrained_All}
\end{center}
\vspace{-0.25cm}
\end{figure}
\begin{figure}
\begin{center}
{\includegraphics[width=\linewidth]{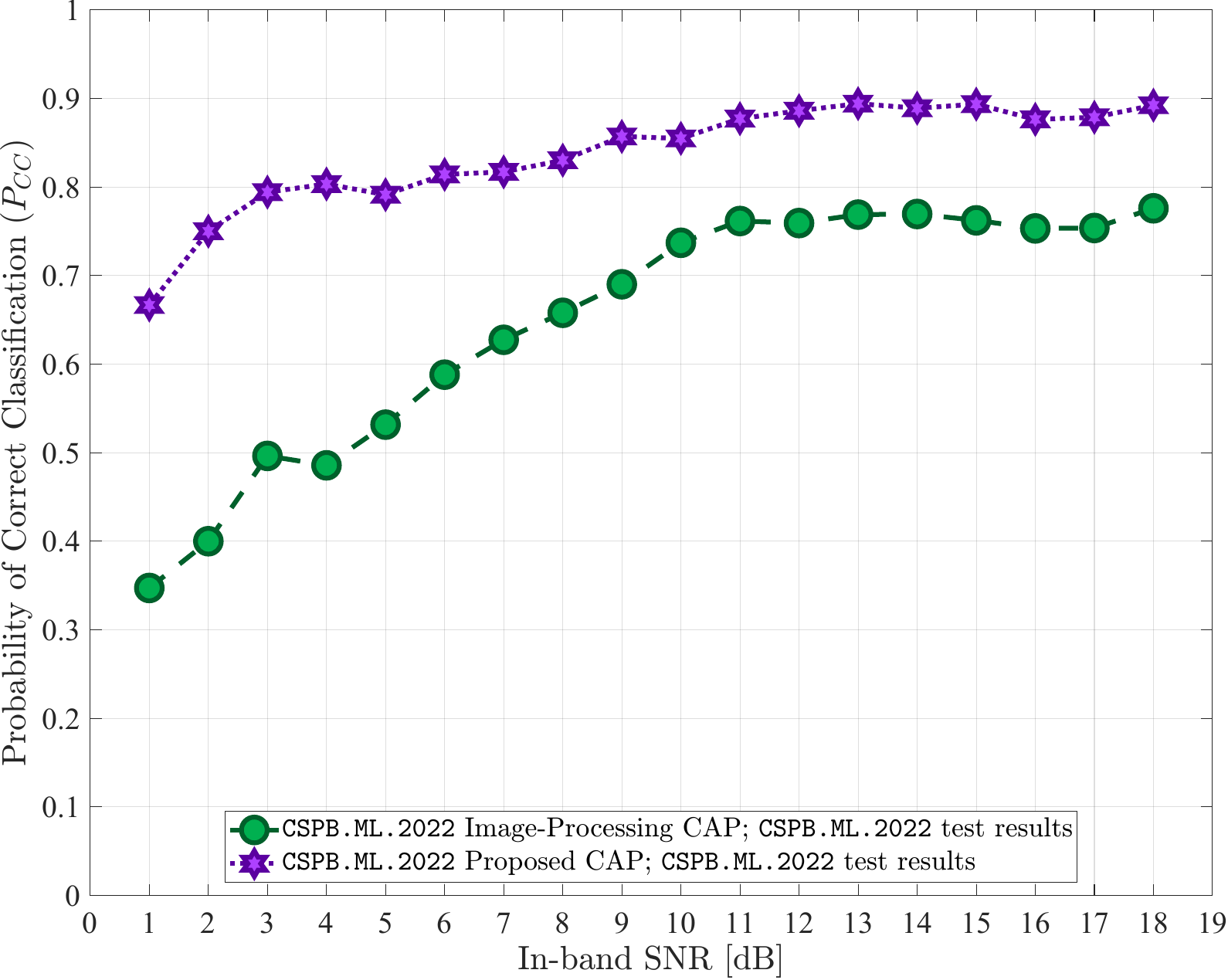}}
\caption{Performance of NNs trained and tested on \texttt{CSPB.ML.2022} dataset.}\label{fig:SNR_GCTest_GCTrained_All}
\end{center}
\end{figure}
\begin{figure}
\begin{center}
{\includegraphics[width=\linewidth]{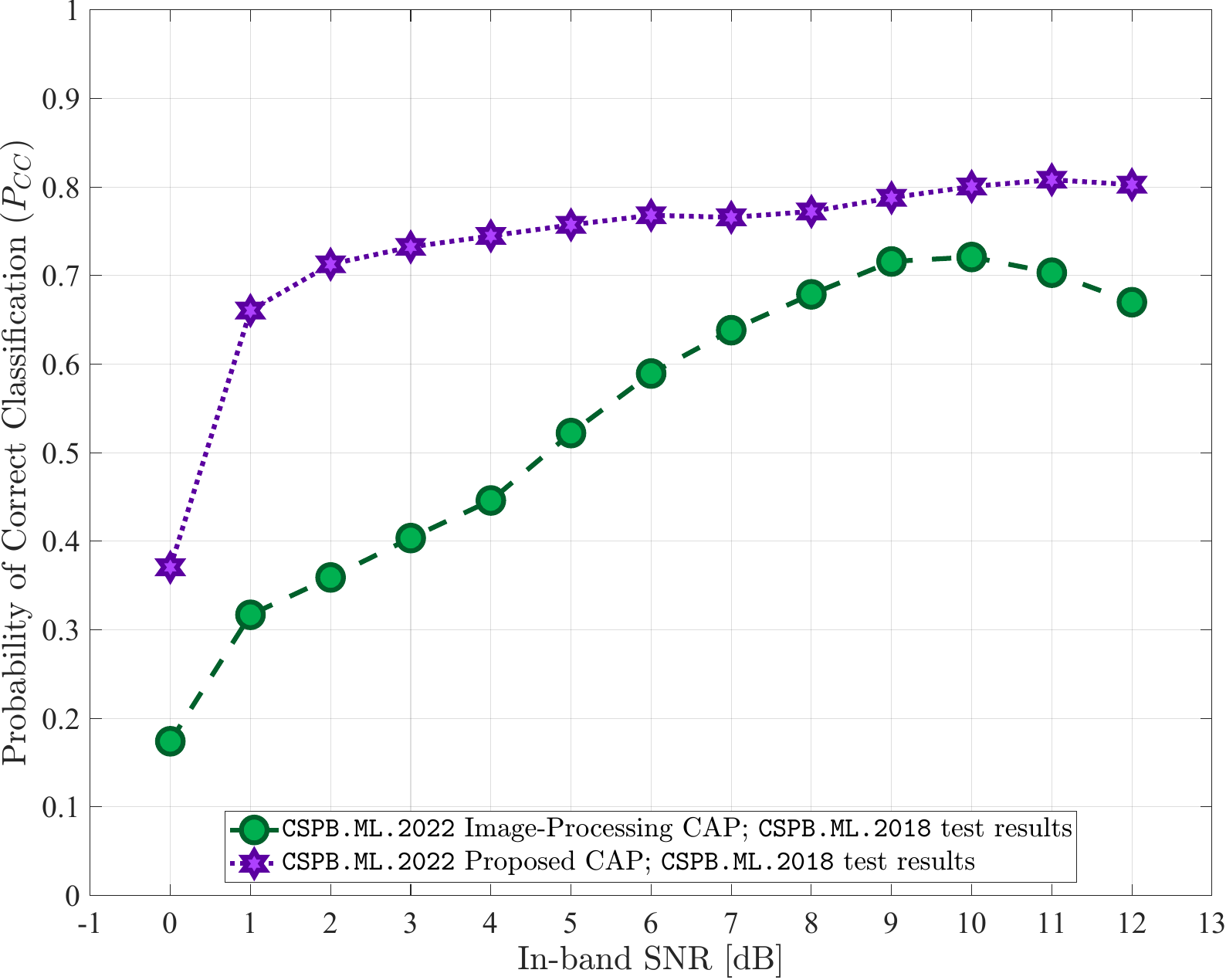}}
\caption{Generalization test results for NNs trained on dataset \texttt{CSPB.ML.2022} and
tested on \texttt{CSPB.ML.2018}.}\label{fig:SNR_CTest_GCTrained_All}
\end{center}
\vspace{-0.25cm}
\end{figure}

\begin{itemize}
\item In the first training instance, dataset \texttt{CSPB.ML.2018} was used, splitting the available signals into $70\%$ for training, $5\%$ for validation,
and $25\%$ for testing. The probability of correct classification for the test results obtained using the $25\%$ test portion of signals in \texttt{CSPB.ML.2018}
is shown in Fig.~\ref{fig:SNR_CTest_CTrained_All}.

\hspace{0.25cm} The CAPs trained on \texttt{CSPB.ML.2018} are then tested on dataset \texttt{CSPB.ML.2022} to assess the generalization abilities
of a trained CAP when classifying all signals available in \texttt{CSPB.ML.2022}. The probability of correct classification for this test is shown
in Fig.~\ref{fig:SNR_GCTest_CTrained_All}.
\item In the second training instance, the NN was reset and trained anew using signals in dataset \texttt{CSPB.ML.2022}, with a similar split of
$70\%$ signals used for training, $5\%$ for validation, and $25\%$ for testing. The probability of correct classification for the test results obtained
using the $25\%$ test portion of signals in \texttt{CSPB.ML.2022} is shown in Fig.~\ref{fig:SNR_GCTest_GCTrained_All}.

\hspace{0.25cm} The CAPs trained on \texttt{CSPB.ML.2022} are then tested on dataset \texttt{CSPB.ML.2018} to assess the generalization abilities
of the trained NN when classifying all signals available in \texttt{CSPB.ML.2018}. The probability of correct classification for this test is shown
in Fig.~\ref{fig:SNR_CTest_GCTrained_All}.
\end{itemize}

In summary, from the plots shown in Figs.~\ref{fig:SNR_CTest_CTrained_All}~--~\ref{fig:SNR_CTest_GCTrained_All}, we note that the proposed
CAP with custom feature extraction layers outperforms the alternative CAP from~\cite{Latshaw_COMM2022} in all training and testing scenarios,
displaying high classification performance and generalization abilities.

The overall probability of correct classification ($P_{CC}$) for all experiments performed is shown in Table~\ref{table:OverallPerformance} and
provides further confirmation that the proposed CAP containing the novel custom layers is able to perform feature extraction tailored to the modulated
signals of interest more effectively than the alternative CAP in~\cite{Latshaw_COMM2022}, which employs only the conventional NN layers used in
the context of image processing. The lowest classification performance seen by the proposed CAP came from training on \texttt{CSPB.ML.2022} and
testing on \texttt{CSPB.ML.2018} with an overall correct classification probability $P_{CC} = 77.1\%$, while the highest classification performance
observed for the alternative CAP in~\cite{Latshaw_COMM2022} came from training on \texttt{CSPB.ML.2022} and testing on \texttt{CSPB.ML.2022},
resulting in a corresponding overall $P_{CC} = 72.1\%$. Thus, even the lowest performance of the proposed CAP exceeds by $5\%$  the best
performance of the alternative CAP in~\cite{Latshaw_COMM2022}.

\begin{figure}
\begin{center}
{\includegraphics[width=\linewidth]{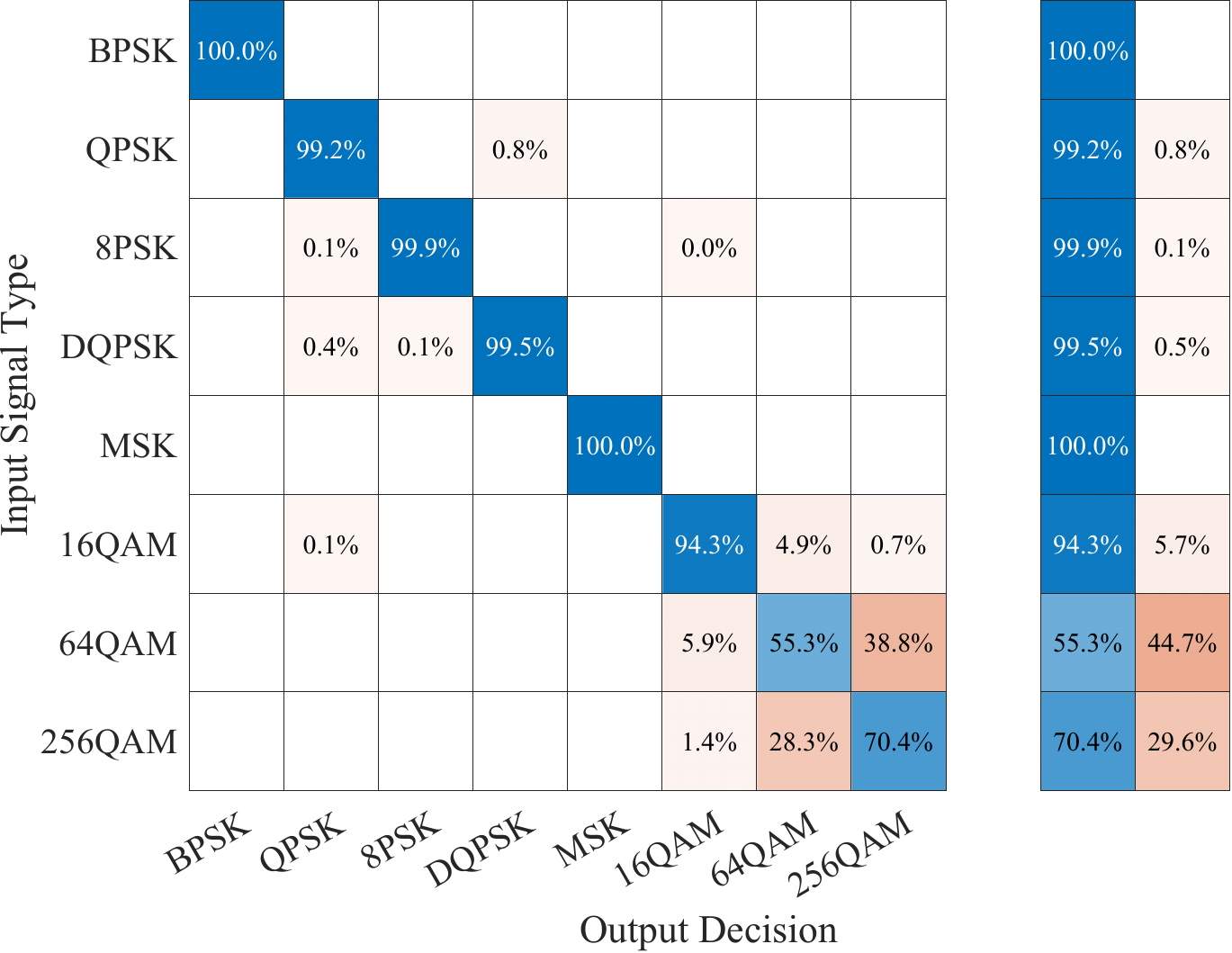}}
\caption{Confusion matrix results for the proposed CAP trained and tested on dataset \texttt{CSPB.ML.2018}.}\label{fig:CM_CTest_CTrained}
\end{center}
\vspace{-0.5cm}
\end{figure}

\begin{table}
\centering
\caption{Classification Performance}
{
\vspace{-0.25cm}
\begin{tabular}{ c c c }
\hline\hline
Classification Model & \begin{tabular}{@{}c@{}} \texttt{CSPB.ML.2018} \\ Test Results\end{tabular} & \begin{tabular}{@{}c@{}} \texttt{CSPB.ML.2022} \\ Test Results\end{tabular} \\
\hline
\begin{tabular}{@{}c@{}} \texttt{CSPB.ML.2018} \\ Alternative CAP~\cite{Latshaw_COMM2022} \end{tabular} & $67.7\%$ & $68.0\%$ \\
\begin{tabular}{@{}c@{}} \texttt{CSPB.ML.2022} \\ Alternative CAP~\cite{Latshaw_COMM2022} \end{tabular} & $62.5\%$ & $72.1\%$ \\
\\
\begin{tabular}{@{}c@{}} \texttt{CSPB.ML.2018} \\ Proposed CAP\end{tabular} & $89.8\%$ & $78.2\%$ \\
\begin{tabular}{@{}c@{}} \texttt{CSPB.ML.2022} \\ Proposed CAP\end{tabular} & $77.1\%$ & $86.7\%$ \\
\hline\hline
\end{tabular}
}\label{table:OverallPerformance}
\vspace{-0.25cm}
\end{table}

\begin{figure}
\begin{center}
{\includegraphics[width=\linewidth]{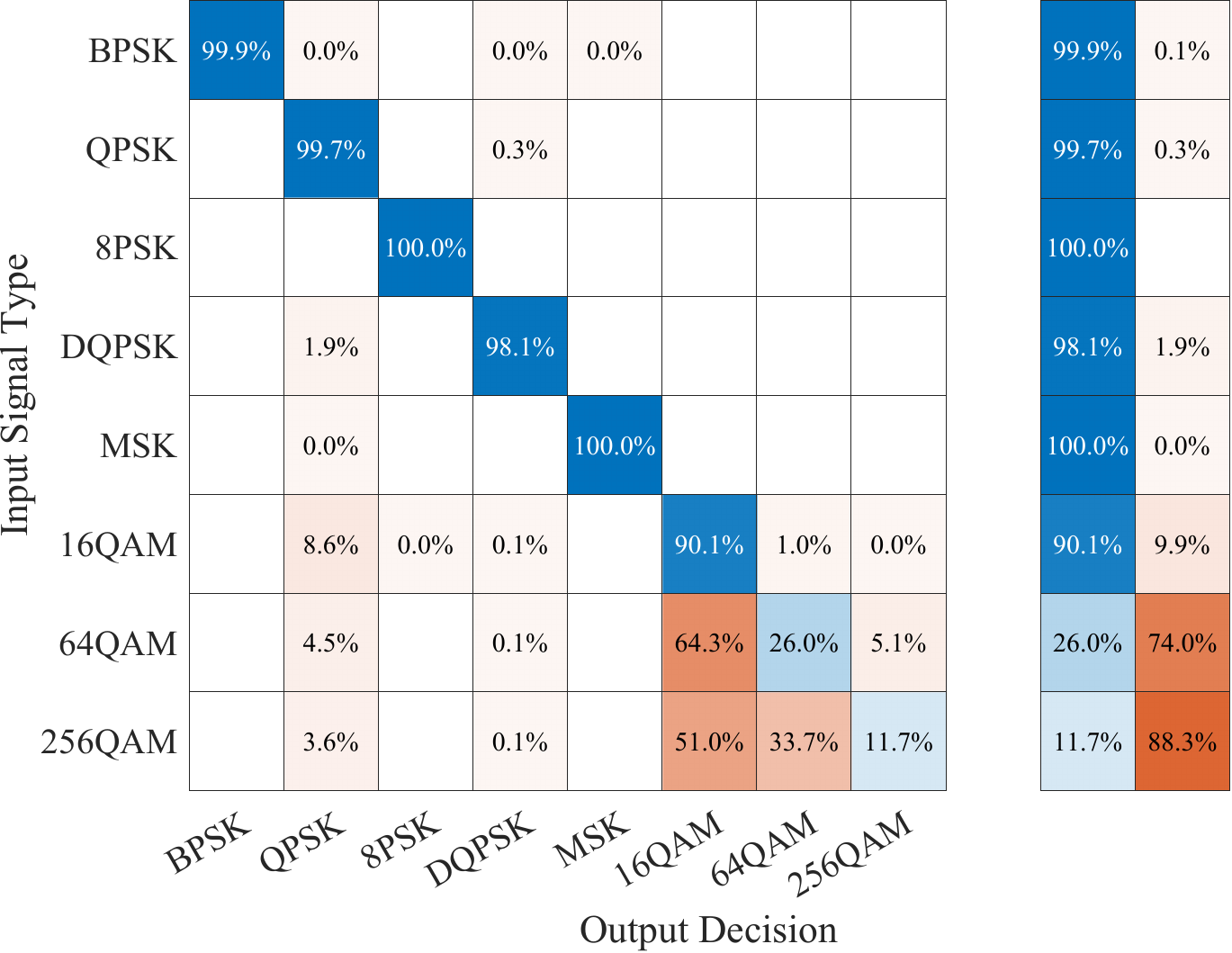}}
\caption{Confusion matrix generalization results for the proposed CAP trained on dataset \texttt{CSPB.ML.2018} and tested on dataset \texttt{CSPB.ML.2022}.}\label{fig:CM_GCTest_CTrained}
\end{center}
\vspace{-0.5cm}
\end{figure}

For additional insight into the classification performance of the proposed CAP with custom feature extracting layers, we have also looked at its corresponding
confusion matrices, which are shown in Figs.~\ref{fig:CM_CTest_CTrained}~and~\ref{fig:CM_GCTest_CTrained}. As can be observed from these figures,
the classification performance for individual BPSK, QPSK, 8-PSK, $\pi/4$-DQPSK, and MSK modulation schemes when the CAP is trained and tested on
\texttt{CSPB.ML.2018} dataset is within $1.5\%$ difference of the performance observed when the CAP is trained on \texttt{CSPB.ML.2018} dataset
and tested on \texttt{CSPB.ML.2022} dataset\footnote{Similar confusion matrices are obtained for the cases when the CAP is trained and tested on
\texttt{CSPB.ML.2022} dataset and when the CAP is trained on \texttt{CSPB.ML.2022} dataset and tested on \texttt{CSPB.ML.2018}, but the corresponding
plots are omitted due to space considerations.}, indicating excellent generalization performance for these five types of digital modulation schemes.
Furthermore, the only modulation schemes for which the classification performance of the proposed CAP decreased when the testing dataset was
changed from \texttt{CSPB.ML.2018} to \texttt{CSPB.ML.2022} are 16-QAM, 64-QAM, and 256-QAM.  

The reason the proposed CAP is unable to generalize its training well for the three QAM modulation types is due to the fact that these modulation
types are only distinguishable from each other at higher orders when their signal power is known or accurately estimated. Since we normalized the
total (signal+noise) power rather than just the signal power, the relative differences between the power levels of the 16-QAM, 64-QAM, and 256-QAM
higher-order moments were removed. Thus, the differences between the extracted features for 16-QAM, 64-QAM, and 256-QAM were different enough
between \texttt{CSPB.ML.2018} and \texttt{CSPB.ML.2022} that the proposed CAP could not obtain good generalization on these three modulation types.

In future work, we plan to investigate additional custom NN function layers that process the I/Q signal data with only signal power normalization, rather
than total power normalization, in order to overcome the issue of poor generalization for QAM modulation types, without needing to resort to full estimates
of the CCs.

\section{Conclusions}\label{sec:Conclusion}
This paper presents a novel DL-based neural network (NN) classifier for digitally 
modulated signals that uses a capsule (CAP) network with custom feature-extraction layers.
The proposed CAP takes as input pre-processed I/Q data, on which blind band-of-interest
(BOI) estimation has been applied to enable
out-of-band noise filtering and CFO-estimate-based spectral centering.
The input data is normalized to unit total power prior to input
to the CAP. The proposed custom function layers are essentially
homogeneous even-order nonlinear functions, and perform feature 
generation and extraction on the pre-processed I/Q data more reliably and
predictably than convolutional NN layers alone, resulting in very good 
classification and generalization performance. Prior to this effort, we have
been unable to achieve significant generalization using any conventional
CAP or convolutional NN that takes I/Q data as input and know of no other 
published successes. 

Future research will involve additional novel NN function layers that enable the 
use of I/Q data which has been normalized to unit signal power rather
than unit total (signal+noise) power, to further improve classification and 
generalization performance for QAM modulation schemes of interest
(16-QAM, 64-QAM, and 256-QAM). Additionally, the inclusion of NN function layers that 
extract features proportional to non-conjugate CCs ($n = 2m$)
and conjugate CCs, where the number of conjugations is greater than zero 
($n \neq 2m$ and $m \neq 0$), will be considered as this may prove necessary
to distinguish between expanded input classes, such as multiple kinds of FSK/CPM/CPFSK, 
which can have little or no cyclostationarity for $n \neq 2m$. In this way we
intend to create neural-network classifiers that can accept I/Q data samples
as input, yet simultaneously produce excellent classification performance
and high generalization, proving to be highly robust in the face of probability-model
departures from the training-set density functions.

\section*{Acknowledgment}
The authors would like to acknowledge the use of Old Dominion University High-Performance Computing facilities for obtaining numerical results
presented in this work.



 \bibliographystyle{IEEEtran}
 \bibliography{./bib/refs}

%
%
%

\end{document}